\documentclass[apjl]{emulateapj}
\usepackage{comment}
\usepackage[T1]{fontenc}
\usepackage{pslatex}

\shorttitle{STAR FORMATION AND THE FIR-RADIO CORRELATION WITHIN GALAXIES}
\shortauthors{MURPHY ET AL.}

\submitted{Accepted into \apjl~Letters, September 26, 2006}
\journalinfo{Accepted into \apjl~Letters, September 26, 2006}

\begin{document}

\title{The Effect of Star Formation on the Far-Infrared--Radio
  Correlation within Galaxies} 
\author{
E.J.~Murphy,\altaffilmark{1} G.~Helou,\altaffilmark{2}
R.~Braun,\altaffilmark{3} J.D.P.~Kenney,\altaffilmark{1}
L.~Armus,\altaffilmark{2} D.~Calzetti,\altaffilmark{4}
B.T.~Draine,\altaffilmark{5} R.C.~Kennicutt, Jr.,\altaffilmark{6,7}
H.~Roussel,\altaffilmark{8} F.~Walter,\altaffilmark{8}
G.J.~Bendo,\altaffilmark{9} B.~Buckalew,\altaffilmark{2}
D.A.~Dale,\altaffilmark{10} C.W.~Engelbracht,\altaffilmark{7}
J.D.T.~Smith,\altaffilmark{7} M.D.~Thornley\altaffilmark{11}
}
\altaffiltext{1}{\scriptsize Department of Astronomy, Yale University,
  P.O. Box 208101, New Haven, CT 06520-8101; murphy@astro.yale.edu}
\altaffiltext{2}{\scriptsize California Institute of Technology,
  Pasadena, CA 91101}
\altaffiltext{3}{\scriptsize ASTRON, 7990 AA Dwingeloo, The
  Netherlands} 
\altaffiltext{4}{\scriptsize Space Telescope Science Institute,
  Baltimore, MD 21218}
\altaffiltext{5}{\scriptsize Princeton University Observatory,
  Princeton, NJ 08544} 
\altaffiltext{6}{\scriptsize Institute of Astronomy, University of
  Cambridge, Cambridge CB3 0HA, UK}
\altaffiltext{7}{\scriptsize Steward Observatory, University of
  Arizona, Tucson, AZ 85721}
\altaffiltext{8}{\scriptsize Max Planck Institut f\"{u}r
  Astronomie,D-69117 Heidelberg, Germany}  
\altaffiltext{9}{\scriptsize Imperial College, London SW7 2AZ UK}
\altaffiltext{10}{\scriptsize University of Wyoming, Laramie, WY 82071}
\altaffiltext{11}{\scriptsize Bucknell University, Lewisburg, PA 17837}

\begin{abstract}
Using data obtained for twelve galaxies as part of the {\it Spitzer}
Infrared Nearby Galaxies Survey (SINGS) and the Westerbork Synthesis
Radio Telescope (WSRT)-SINGS radio continuum survey, we study how star
formation activity affects the far-infrared (FIR)--radio correlation
{\it within} galaxies by testing a phenomenological model 
which describes the radio image as a smeared version of the FIR
image.
The physical basis of this description is that cosmic-ray (CR)
electrons will diffuse measurably farther than the mean free path of
dust-heating photons before decaying by synchrotron radiation.  
This description works well in general.  
Galaxies with higher infrared surface brightnesses 
have best-fit smoothing scale-lengths of a few hundred parsecs,
substantially shorter than those for lower surface brightness
galaxies.
We interpret this result to suggest that galaxies with higher disk
averaged star formation rates have had a recent episode of enhanced
star formation 
and are characterized by a higher fraction of young CR electrons that
have traveled only a few hundred parsecs from their acceleration sites
in supernova remnants 
compared to galaxies with lower star formation activity.
\end{abstract}
\keywords{cosmic rays --- infrared: galaxies --- radio
  continuum: galaxies}

\section{Introduction}
Radio continuum emission from normal galaxies arises from a
combination of non-thermal (synchrotron) and thermal (free-free)
processes; 
the former critically depends on a galaxy's magnetic field and cosmic
ray (CR) electron distributions.
To date, the evolution of CRs has been studied directly only in the
Milky Way. 
In other galaxies, information has come from sampling the CR electron
component via multi-frequency radio observations
\citep[e.g.][]{nd91,dlg95,ul96,ji99,rb05}.
Radio continuum data alone, however, only describe the present
distribution of CR electrons, without providing any information
regarding the initial source distribution or propagation history.

The close spatial correlation between the thermal dust and non-thermal
synchrotron emission within galaxies argues for a shared origin,
presumably massive star formation \citep{har75,gxh85}.
Young massive stars, still in or near their dusty natal molecular cloud,
emit photons that are re-radiated by dust grains at far-infrared (FIR)
wavelengths.  
These stars are also the progenitors of supernovae (SNe)
whose remnants (SNRs) are thought to accelerate CR electrons
responsible for synchrotron radiation.
Since the mean free path of dust-heating photons ($\sim$100~pc) is far
shorter than the diffusion length of CR electrons ($\sim$1-2~kpc),
\citet{bh90} conjectured that the radio image of a galaxy should
resemble a smoothed version of its infrared image. 
This prescription has been shown to hold for galaxies observed at the
``super resolution'' ($\la$1$\arcmin$) of {\it IRAS} HIRES data
\citep{mh98}.  
Using 
{\it Spitzer} imaging, \citet[hereafter M06]{ejm06} tested this model
on sub-kiloparsec scales within four nearby galaxies and found it
applicable to first order, improving the correlation between the
infrared and radio maps while leaving patterns in the residuals that
mirror the structure in the galaxy's image.   
More recently, \citet{ah06} has found that within the Large Magellanic
Cloud (LMC), synchrotron halos around individual star-forming regions
are more extended than FIR-emitting regions, 
corroborating this phenomenology on scales $\ga$50~pc.

In this {\it Letter} we explore how star formation activity affects
the relative spatial distributions of FIR and radio continuum emission
for a sample of 12 spiral galaxies observed as part of the {\it
  Spitzer} Infrared Nearby Galaxies Survey (SINGS) legacy science
project \citep{rk03} and the Westerbork Synthesis Radio Telescope
(WSRT)-SINGS radio continuum survey.   
This analysis builds on that of M06, though the larger sample has
increased the range in star formation rates (SFRs) investigated by a
factor of $\sim$5.
Such a study is only now possible due to the high spatial resolution
and sensitivity of the {\it Spitzer} Space Telescope at FIR
wavelengths.

%

\section{Data and Analysis}
\subsection{Observations and Data Reduction}
Observations at 24, 70, and 160~$\micron$ were obtained using the
Multiband Imaging Photometer for {\it Spitzer}
\citep[MIPS;][]{gr04} as part of the SINGS legacy science
program. 
The MIPS data were processed using the MIPS Data Analysis Tool
\citep[DAT;][]{kdg05} and included in the SINGS data release 3 (DR3).
Details on the data reduction can be found in M06.
The full width at half maximum (FWHM) of the MIPS 24, 70, and
160~$\micron$ point spread functions (PSFs) are 5$\farcs$7, 17$\arcsec$,
and 38$\arcsec$, respectively, and the final calibration uncertainty is
$\sim$10~\% at 24~$\micron$ and $\sim$20~\% at 70 and 160~$\micron$.

Radio continuum imaging at 22~cm was performed using the WSRT.
Each galaxy was observed in a configuration with particularly good
sampling at short baselines to  
curtail flux loss at low spatial frequencies.
The WSRT data were CLEANed and self-calibrated using an imaging
pipeline based on the MIRIAD package.  
The final CLEAN maps were restored with 18$\arcsec$ FWHM circular
Gaussian beams.
Total intensity calibration of the data was performed in the AIPS
package and the flux density calibration of the radio maps is better
than 5~\%. 
A complete description of the image processing can be found in
\citet{rb06}.  

To properly match the resolution of the MIPS and radio images, the
Fourier Transform (FT) of the 70~$\micron$ images was divided by
the FT of a model of the MIPS 70~$\micron$ PSF, and then multiplied by
the FT of a Gaussian PSF matching that used as the CLEAN restoring
beam.
The final product was then transformed back into the image plane, and
checked to ensure flux was conserved.
By matching the resolution of the MIPS and radio data with Gaussian
beams, rather than the MIPS 70~$\micron$ PSF (which has significant
power in its wings), the dynamic range of the 70~$\micron$ images has
increased by roughly a factor $\sim$2.
Due to the improved resolution there is also a slight increase in the
RMS noise of the maps by an average of $\sim$33~\%.


\subsection{Image-Smearing Model}\label{secres}
We smooth each galaxy's 70~$\micron$ map with a parameterized kernel
and calculate the normalized squared residuals between the smoothed
70~$\micron$ and the 22~cm image as defined by,
\begin{equation}
\phi(Q,l) = \frac{\sum[Q^{-1}\tilde{I}_{j}(l) - R_{j}]^{2}}{\sum
  R_{j}^2}, 
\label{phi}
\end{equation} 
where \(Q=\sum\tilde{I}_{j}/\sum R_{j}\) is
used as a normalization factor, $\tilde{I}({l})$ represents the
infrared image after smearing with kernel of scale-length $l$,
$R$ is the observed radio image, and the subscript $j$ indexes
each pixel. 
The smoothing kernel takes the form, \(\kappa(r) = e^{-r/r_{\rm o}}\),
where $r_{0}$ is the projected, $e$-folding length of the kernel
($l$).
This kernel was chosen because it 
works as well as or better than the other kernel types studied by
M06.

We calculate the residuals after first removing pixels detected
below the 3~$\sigma$ level in either the radio or maximally-smeared 
infrared maps and editing out contaminating background radio
sources. 
Additional editing was performed in the cases of NGC~3031 and
NGC~5194 as described in M06.

\begin{figure}
\resizebox{8.8cm}{!}{
\plotone{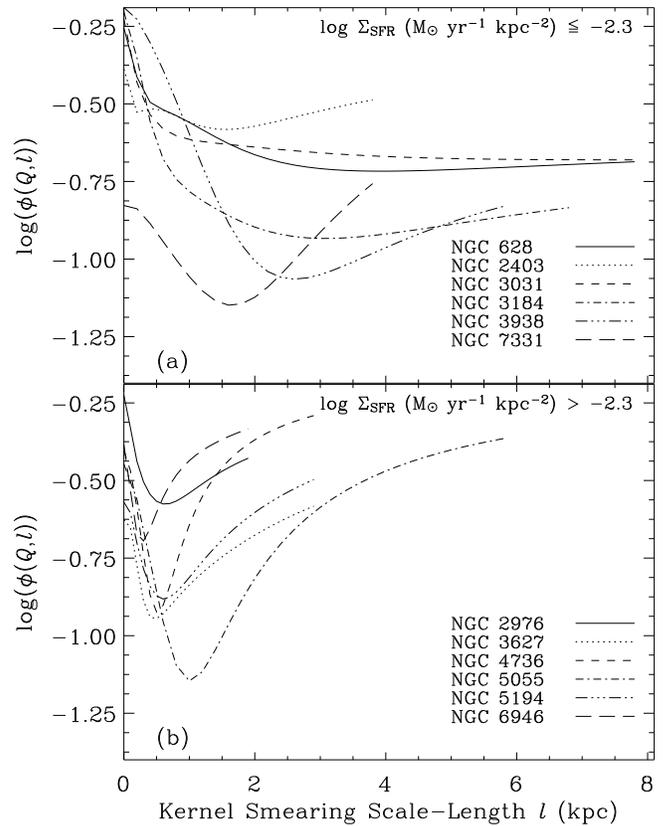}} 
\caption{Residuals between the observed radio maps and smeared
    70~$\micron$ images (as defined in $\S$\ref{secres}) as a function of
    smearing scale-length.  
    Galaxies having low disk-averaged star formation rates, defined
    by \(\log~\Sigma_{\rm SFR} \leq -2.3\) (see $\S$\ref{secsfrd}) are
    plotted in panel (a) while galaxies with high star formation
    activity, and larger values of $\Sigma_{\rm SFR}$, are plotted in
    panel (b).
    \label{rescurves}}
\end{figure}

The residuals are plotted against the smearing kernel scale-length in
Figure \ref{rescurves}.
To quantify the improvement in the spatial correlation between the
70~$\micron$ and 22~cm maps due to smoothing the 70~$\micron$ image,
we measure $\Phi$ defined as the logarithmic decrement from
$\phi(Q,0)$ to $\phi(Q,l)$ at its minimum; this minimum point defines
the best-fit scale length.
Since $\Phi$ is an estimate of the {\it global} improvement, we also
created residual maps, defined by 
\(\log~(Q^{-1}\tilde{I}) - \log~R\), 
using the best-fit smearing kernels to determine how well the smearing
prescription works for different regions within each galaxy.
The 70~$\micron$ and residual maps are plotted in Figure \ref{resmaps}.

\subsection{SFR Surface Densities}\label{secsfrd}
To see how the FIR-radio correlation within galaxies is affected by
ongoing star formation activity, we measure their total infrared (TIR;
3-1100~$\micron$) surface brightness and express it as SFR surface 
density, $\Sigma_{\rm SFR}$.
We first create $I_{\rm TIR}$ maps 
via a weighted combination of the 3 MIPS maps according to Equation 4
of \citet{dd02}.
The SFR surface densities, 
in units M$_{\sun}$~yr$^{-1}$~kpc$^{-2}$, were then calculated by
estimating a SFR \citep[i.e. Equation 5 of][]{eb03} using the TIR
luminosity 
within an ellipse fit to 
the $1.4\times10^{-7}$~W~m$^{-2}$~sr$^{-1}$ isophotal radius, and then
dividing by the corresponding deprojected area (comparable to the
galaxy sizes at 70~$\micron$).

\section{Results}
In Figure \ref{rescurves}, we plot the residuals against the smearing
scale-length for the sample galaxies, separated into low and high
$\Sigma_{\rm SFR}$ bins.
There is a wide range of behavior in the residual curves among the
galaxies.
Three galaxies do not display any definitive minimum in their curves.
We also find that galaxies with higher $\Sigma_{\rm SFR}$ have better
defined single minima compared to galaxies with lower $\Sigma_{\rm
  SFR}$ even though the distribution of $\Phi$ is similar between the
two groups.

The smearing technique improves the correlation between
the radio and 70~$\micron$ maps by an average of $\sim$0.5 in
$\log~\phi$.   
This value is 0.3~dex larger than what was found by M06 and is due to
improvements in the resolution matching and the larger sample. 

We display the 70~$\micron$ and residual maps of each galaxy ordered
by 
$\Sigma_{\rm SFR}$ in Figure \ref{resmaps}.
While there 
appears to be a general correlation between the
structure found in the 70~$\micron$ and residual maps, there is
a range in the residual map behavior exemplified by the cases of
NGC~628, NGC~5055, and NGC~6946.
For NGC~5055, 
the residuals are small and the image is nearly uniform. 
In NGC~6946, on the other hand, we see distinct structures in the
residual maps associated with the spiral-arm morphology of the galaxy,
as well as a much larger range in the residuals 
compared to NGC~5055.
The arms of NGC~6946 display infrared excesses peaked on star-forming
regions while the galaxy's inter-arm regions exhibit excess radio
emission.
NGC~628 is distinctly opposite to NGC~6946 in its residual map
displaying {\it radio} excesses for star-forming regions.
This implies that the large smoothing scale-lengths, which provide the
best-fits for low $\Sigma_{\rm SFR}$ galaxies, oversmooth the active 
star forming components.

In Figure \ref{SFRDleg} we plot the best-fit scale-length for each
galaxy versus their SFR surface density, $\Sigma_{\rm SFR}$.
Galaxies are better modeled by
larger smearing scale-lengths if their disk-averaged SFR is lower.
To ensure that this trend is not an artifact, we have compared the
best-fit scale-lengths with galaxy distances, linear sizes, and
signal-to-noise (S/N) ratios and find a correlation with none of them.
The implications of this trend will be discussed in
$\S$\ref{secdisc}. 

\subsection{Effects of Free-Free Emission}\label{sectrad}
Since the image-smearing technique is designed for the physics of
non-thermal radio emission, the presence of free-free emission can 
complicate the interpretation of the best-fit smearing scale-lengths.  
To gauge the effect of free-free emission on the trend in Figure
\ref{SFRDleg}, we repeat the smearing analysis 
after subtracting an estimate of the free-free emission distribution
using a scaled version of the 24~$\micron$ maps. 

The scaling factor was determined by first relating the 24~$\micron$
emission to Pa$\alpha$ line emission using the correlation found
between 24~$\micron$ and Pa$\alpha$ luminosities within NGC~5194 (M51a)
\citep{dc05}.  
Although this correlation is not universal \citep{ppg06}, it should
be sufficient for the purpose of roughly estimating how the thermal
emission affects the best-fits scale-lengths.
The corresponding Pa$\alpha$ emission estimate can then be related to
an ionizing photon rate \citep{ost89}, and the expected free-free
emission at 1.4~GHz \citep{rub68}.
The final relation is given by,
\begin{equation}
\left(\frac{S_{\rm T}}{\rm Jy}\right) \sim 7.93\times10^{-3}
\left(\frac{T}{10^{4}~{\rm K}}\right)^{0.45} 
\left(\frac{\nu}{\rm GHz}\right)^{-0.1}
\left(\frac{f_{\nu}(24~\micron)}{\rm Jy}\right),
\end{equation}
where we have assumed an average H~{\small II} region temperature of
\(T=10^{4}\)~K.

For seven of the sample galaxies, \citet{nkw97} report a mean
thermal fraction of $\sim$0.08 at 1.0~GHz from radio spectral index
fitting.
Assuming a mean spectral index of $-0.8$, we obtain a very similar
thermal fraction using our data. 
%
Since the distribution of free-free emission is clumpy, mostly
originating in H~{\small II} regions, we expect that the smearing
scale-lengths should increase after it is removed.
The best-fit scale-lengths increase
by an average of $\sim$25~\%.
The slope of the fit using these scale-lengths versus $\Sigma_{\rm
  SFR}$ is shallower by $\la$10~\%, indicating that the observed
trend in Figure \ref{SFRDleg} cannot be suppressed by removing the
thermal radio emission.

\begin{figure}
\setcounter{figure}{2}
\resizebox{8.8cm}{!}{
\plotone{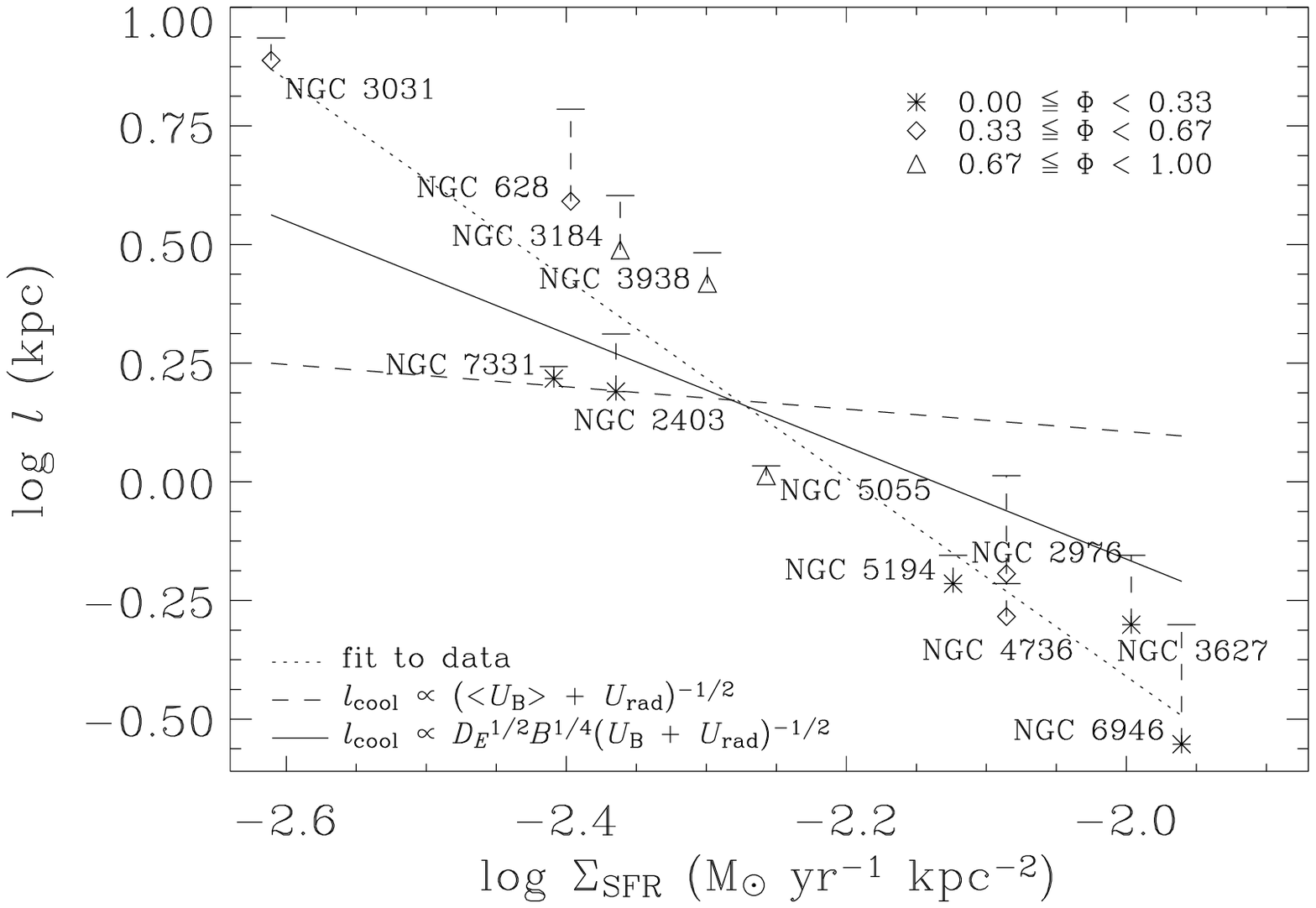}}
\caption{The SFR surface densities, ($\Sigma_{\rm SFR}$, see
    $\S$\ref{secsfrd}), for each galaxy are plotted against the best-fit
    smoothing kernel scale-length.
    The general improvement in the spatial correlation between the
    70~$\micron$ and 22~cm maps ($\Phi$, see $\S$\ref{secres}) is
    indicated by the plotting symbol. 
    Also plotted is the fit to the data ({\it dotted line}) along with
    the expected diffusion scale-lengths due to inverse Compton (IC)
    losses in a fixed magnetic field ({\it dashed line}) and
    synchrotron + IC losses with an energy-dependent diffusion
    coefficient $D_{E}$ for the steepest possible index ({\it solid
    line}) (see $\S$\ref{secdisc}).  
    The increase to the best-fit scale-length using 22~cm maps
    corrected for free-free emission is indicated by a vertical line
    for each galaxy (see $\S$\ref{sectrad}).
     \label{SFRDleg}}  
\end{figure}

\section{Discussion}\label{secdisc}
We have analyzed the effect of star formation intensity
on the FIR-radio correlation {\it within} galaxies using a
phenomenological image-smearing model. 
We find that the best-fit scale-length decreases as the star
formation activity within each galaxy increases, implying a systematic
decrease in  the mean distance diffused by CR electrons \citep{hb93}.
The measured diffusion scale-lengths can provide independent estimates
for the ages of related star formation episodes as well as help
constrain interstellar medium (ISM) parameters associated with the
propagation characteristics of CR electrons.  

The physical interpretation of our results indicates that CR electrons
are, on average, closer to their place of origin in galaxies having
higher star formation activity. 
Since the diffusion scale-length of CR electrons depends only on their
age and ability to diffuse through a galaxy's ISM, three 
explanations of the results are possible.  The CR electrons may:
(1) have relatively short lifetimes due to a high energy loss
rate;
(2) diffuse at a slower rate due to the ISM having a high density
and magnetic field strength, resulting in a shorter mean free path; or 
(3) have been accelerated recently and be relatively
young.
The first two of these explanations are applicable in the case of
steady-state star formation, while the third requires a recent episode
of enhanced star formation.
We will try to distinguish among these three explanations.

As CR electrons propagate through the ISM, they lose energy due to
synchrotron, inverse-Compton (IC) scattering, bremsstrahlung,
ionization, and adiabatic expansion losses. 
In normal star-forming galaxies only the first two mechanisms are
significant for CR electrons associated with synchrotron emission at
$\sim$1~GHz \citep{jc92}.
Assuming an isotropic distribution of electron
velocities,\footnote[1]{For isotropically distributed electron
  velocities we assume \(<\sin^{2}\theta> = \twothirds\), where
  $\theta$ is the CR electron pitch angle.} 
the cooling timescale due to synchrotron and IC losses for CR
electrons emitting at a frequency $\nu$, in a magnetic field of
strength $B$ is,
\begin{equation}
  \left(\frac{\tau_{\rm cool}}{\rm yr}\right) \sim 2.8\times10^{7}
  \left(\frac{\nu}{\rm GHz}\right) ^{-1/2}
  \left(\frac{B}{\rm \mu G}\right)^{1/2}
  \left(\frac{U_{\rm B}+U_{\rm rad}}{10^{-12}~{\rm
  erg~cm^{-3}}}\right)^{-1}, 
\end{equation} 
where $U_{\rm B}$ and $U_{\rm rad}$ are the magnetic
and radiation field energy densities, respectively.

In simple diffusion models, the propagation of CR electrons is
usually characterized by an empirical energy-dependent diffusion
coefficient, $D_{E}$, where \(D_{E} = 10^{29}\)~cm$^{2}$~s$^{-1}$ for
$E < 1$~GeV and \(10^{29}(E/{\rm GeV})^{1/2}\)~cm$^{2}$~s$^{-1}$ for
$E \geq 1$~GeV \citep{ginz80}.  
Though derived for CR nuclei, the above values appear consistent with
empirically measured diffusion coefficients for CR electrons
\citep[e.g.][]{dlg95}.
According to synchrotron theory, CR electrons will emit at \(\nu
\propto B E^{2}\) thereby allowing $D_{E}$ to be expressed as a
function of $B$ for a fixed $\nu$.
Now, assuming that CR electrons diffuse a distance \(l_{\rm cool} =
\sqrt{D_{E}\tau_{\rm cool}}\), we can estimate the relative importance
of IC losses, ISM density ($n$), and $B$ on the distance CR electrons
travel.

Fixing $B$ and scaling \(U_{\rm rad} \propto \Sigma_{\rm SFR}\), we
plot the expected relation between $l_{\rm cool}$ and $U_{\rm rad}$ in
Figure \ref{SFRDleg} (dashed line).
This relation has a slope $\sim$9 times shallower than a
line fit to the data (dotted line).  
It is more likely that $n$, $B$, and $U_{\rm rad}$ will all scale
together, with a $D_{E}$ dependence on $n$.
Accordingly, we introduce the scaling relations,
\begin{equation}
B \propto n^{\beta},~D_{E} \propto n^{\delta}B^{-1/4}, \nonumber
\end{equation} 
where \(\onethird \leq \beta \leq \twothirds\) and \(-1 \leq \delta
\leq -\onethird\) are realistic index ranges \citep{hb93}.

Setting \(U_{\rm B} = U_{\rm rad}\), which has been shown to be valid
for a large sample of spirals \citep{lvx96}, leads to \(B \propto
\Sigma_{\rm SFR}^{1/2}\), \(\tau_{\rm cool} \propto B^{-3/2}\), 
and the dependence 
\begin{equation}
l_{\rm cool} \propto \Sigma_{\rm SFR}^{(\delta/(4\beta) - 7/16)}.
\end{equation}
For the unlikely case that each galaxy takes the extreme values of
\(\beta = \onethird\) and \(\delta = -1\), we obtain the steepest
possible index of $-19/16$ (solid line in Figure \ref{SFRDleg}), which
is still a factor of $\sim$2 times shallower than the line fit to the
data.  
Consequently, differences in the diffusion properties and/or cooling
timescales of each galaxy's CR electron population alone (i.e. the
case of steady-state star formation), cannot explain the observed
trend in Figure \ref{SFRDleg}.

On the other hand, by setting the best-fit scale-lengths equal to 
$l_{\rm cool}$ and $B = 5~\mu$G for each galaxy, we estimate mean ages
ranging from $\sim$1.1$\times 10^{5}$ to $\sim$8.3$\times 10^{7}$~yr for 
these 1.4~GHz emitting populations of CR electrons.
Even if we assume \(B \propto \Sigma_{\rm SFR}^{1/2}\), this range of
ages shrinks only slightly.  
While these ages are plausible, spanning nearly 3 orders of magnitude
over a factor of 4 change in $\Sigma_{\rm SFR}$ seems unphysically
large.
This suggests that as $\Sigma_{\rm SFR}$ decreases the best-fit
scale-length shifts from being dominated by young CR electrons
associated with regions of star formation to being dominated by old CR
electrons in the galaxy's underlying disk.
Consequently, we conclude that galaxies with higher $\Sigma_{\rm SFR}$
are likely to have experienced a recent episode of enhanced star
formation compared to galaxies with lower disk-averaged SFRs, 
leading to a larger fraction of relatively young CR electrons within
their disks. 
We should also note that galaxies with higher star formation activity
may also suffer from increased CR electron escape due to instability
driven breaks in magnetic field lines \citep{ep66}, though
characterizing the importance of escape is beyond the scope of this
{\it Letter}.

The galaxies analyzed here exhibit a wide range of ages and
spreading scale-lengths of their CR electrons, going from the
youngest populations, recently injected within star-forming regions
($l \sim$500~pc), to the oldest CR electrons, making up a
galaxy's underlying synchrotron disk ($l \sim$3~kpc).  
This exponential synchrotron disk results from the superposition of
older CR electrons originating from all star-forming regions. 
In spite of an occasionally dominant mode, a given galaxy will contain
a range of ages for CR electrons, so it is no surprise that a single
smoothing kernel cannot describe perfectly the relation between
infrared and radio disks, as evidenced by the recognizable structure in
the residual maps.
NGC~628, NGC~2403, NGC~3031, and NGC~3184
exhibit inflection points in their residual curves which is also
suggestive of multiple CR electron populations, possibly from
time-separated bursts of star formation. 
A more realistic description to account for the range of ages
translates into a multiscale analysis of the galaxy images
to separate the signatures of various CR electron populations.
This will be the subject of a forthcoming paper.

\acknowledgements
We 
thank Pieter van Dokkum for stimulating discussions.
E.J.M. 
acknowledges support for this work provided by
the {\it Spitzer} Science Center Visiting Graduate Student program.
As part of the {\it Spitzer} Space Telescope Legacy Science Program,
support was provided by NASA through Contract Number 1224769 issued
by the Jet Propulsion Laboratory, California Institute of Technology
under NASA contract 1407.

\clearpage

\setcounter{figure}{1}
\begin{figure}[!ht]
\centerline{\hbox{
  \resizebox{15cm}{!}{
    \plotone{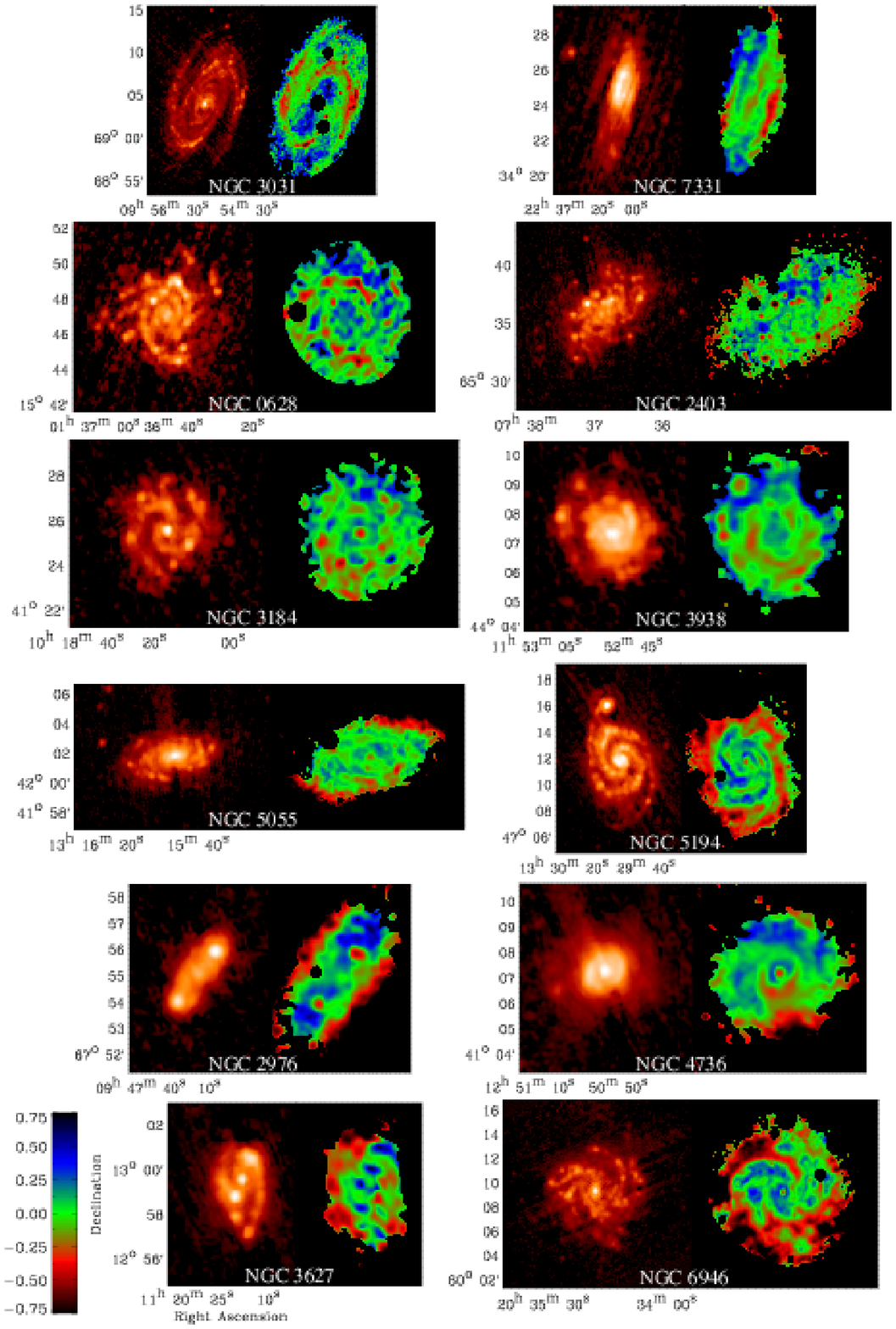}}}}
\caption{The 70~$\micron$ ({\it left}) and residual ({\it right})
    images are plotted for each galaxy and are ordered by increasing
    $\Sigma_{\rm SFR}$ from left to right, top to bottom.  
    The 70~$\micron$ images are logarithmically scaled with a
    stretch ranging from the RMS background level to the maximum
    surface brightness in the galaxy disk.
    Streaking in the 70~$\micron$ images, clearly visible for NGC~4736,
    NGC~5055, and NGC~7331, is a latent image effect that does not
    affect the results (see M06 for details). 
    Residual maps are created by subtracting the observed radio maps
    from the smeared 70~$\micron$ images (as defined in
    $\S$\ref{secres}) for each galaxy using the best-fit exponential
    kernel projected in the plane of its galactic disk.
    Regions removed for the residual calculations (e.g. background
    radio sources) appear as dark, circular holes in a few of the
    residual maps.
    \label{resmaps}}
\end{figure}


\begin{thebibliography}{}
\bibitem[Beck (2005)]{rb05} Beck, R.  2005, in Cosmic Magnetic Fields,
  eds. R. Wielebinsky \& R. Beck (Berlin: Springer), 41 
\bibitem[Bell (2003)]{eb03} Bell, E. F.  2003, \apj, 586, 794
\bibitem[Bicay \& Helou (1990)]{bh90} Bicay M. D. \& Helou, G. 1990,
  \apj, 362, 59
\bibitem[Braun et al. (2006)]{rb06} Braun, R., Oosterloo, T. A.,
  Morganti, R., Klein, U., \& Beck, R. 2006, \aap, submitted
\bibitem[Calzetti et al. (2005)]{dc05} Calzetti, D., et al. 2005,
  \apj, 633, 871  
\bibitem[Condon (1992)]{jc92} Condon, J. J. 1992, \araa, 30, 575
\bibitem[Dahlem et al. (1995)]{dlg95} Dahlem, M., Lisenfeld, U.,
  \& Golla, G.  1995, \apj, 444, 119
\bibitem[Dale \& Helou (2002)]{dd02} Dale, D. A. \& Helou, G. 2002,
  \apj, 576, 159
\bibitem[Duric (1991)]{nd91} Duric N. 1991, in Duric N., Crane P.,
  eds, ASP Conf. Ser. Vol. 18, The Interpretation of Modern Synthesis
  Observations of Spiral Galaxies. Astron. Soc. Pac., San Francisco,
  p. 17 
\bibitem[Ginzburg et al. (1980)]{ginz80} Ginzburg, V. L., Khazan,
  Ya. M., \&  Ptuskin, V. S. 1980, \apss, 68, 295
\bibitem[Gordon et al. (2005)]{kdg05} Gordon, K. D., et al. 2005,
  \pasp, 117, 503 
\bibitem[Harwit \& Pacini (1975)]{har75} Harwit, M, \& Pacini, F
  1975, \apjl, 200, L127
\bibitem[Helou \& Bicay (1993)]{hb93} Helou, G. \& Bicay, M. D. 1993,
  \apj, 415, 93
\bibitem[Helou, Soifer, \& Rowan-Robinson (1985)]{gxh85} Helou, G.,
  Soifer, B. T., \& Rowan-Robinson, M. 1985, \apjl, 298, L7
\bibitem[Hughes et al. (2006)]{ah06} Hughes, A., Wong, T., Ekers, R.,
  Staveley-Smith, L., Filipovic, M., Maddison, S., Fukui, Y., \&
  Mizuno, N. 2006, \mnras, 370, 363
\bibitem[Irwin et al. (1999)]{ji99} Irwin, J. A., English, J., \&
  Sorathia, B. 1999, \aj, 117, 2102 
\bibitem[Kennicutt et al. (2003)]{rk03} Kennicutt, R. C. Jr., et al. 2003,
  \pasp, 115, 928
\bibitem[Lisenfeld et al. (1996a)]{ul96} Lisenfeld, U., Alexander, P.,
  Pooley, G. G., \& Wilding, T. 1996a, \mnras, 281, 301 
\bibitem[Lisenfeld et al. (1996b)]{lvx96} Lisenfeld, U., V\"{o}lk, H. J.,
  \& Xu, C. 1996b, \aap, 306, 677  
\bibitem[Marsh \& Helou (1998)]{mh98} Marsh, K. A. \& Helou, G. 1998,
  \apj, 493, 121
\bibitem[Murphy et al. (2006)]{ejm06} Murphy E. J., et al. 2006, \apj,
  638, 157 (M06)
\bibitem[Niklas et al. (1997)]{nkw97} Niklas, S, Klein, U., \&
  Wielebinski, R. 1997, \aap, 322, 19 
\bibitem[Osterbrock (1989)]{ost89} Osterbrock, D. E. 1989,
  Astrophysics of Gaseous Nebulae \& Active Galactic Nuclei (Mill
  Valley: University Science Books)
\bibitem[Parker (1966)]{ep66} Parker, E. N., 1966, \apj, 145, 811
\bibitem[P\'{e}rez-Gonz\'{a}lez et al. (2006)]{ppg06}
  P\'{e}rez-Gonz\'{a}lez, P. G., et al. 2006, \apj, 648, 987
\bibitem[Rieke, et al. (2004)]{gr04} Rieke, G. H., et al. 2004, \apjs,
  154, 25
\bibitem[Rubin (1968)]{rub68} Rubin, R. H.  1968, \apj, 154, 39
\end{thebibliography}
\end{document}